# Indacenodithiophene homopolymers via direct arylation: direct polycondensation versus polymer analogous reaction pathways

*Desiree Adamczak[1], Hartmut Komber[2], Anna Illy[1], Alberto D. Scaccabarozzi[3], Mario Caironi[3], Michael Sommer[1]\**

[1]Institut fur Chemie, Technische Universitat Chemnitz, Straße der Nationen 62, 09111 Chemnitz, Germany

[2]Leibniz Institut fur Polymerforschung Dresden e. V., Hohe Straße 6, 01069 Dresden, Germany

[3]Center for Nano Science and Technology @PoliMi,Istituto Italiano di Tecnologia,Via Pascoli 70/3, 20133 Milano, Italy

ABSTRACT. Indacenodithiophene (IDT) based materials are emerging high performance conjugated polymers for use in efficient organic photovoltaics and transistors. However, their preparation generally suffers from long reaction sequences and is often accomplished using disadvantageous Stille couplings. Herein, we present detailed synthesis pathways to IDT homopolymers using C-H activation for all C-C coupling steps. Polyketones are first prepared by direct arylation polycondensation (DAP) in quantitative yield and further cyclized polymer analogously. This protocol is suitable for obtaining structurally well-defined IDT homopolymers, provided that the conditions for cyclization are chosen appropriately and that side reactions are suppressed. Moreover, this polymer analogous pathway gives rise to asymmetric side chain patterns, which allows to fine tune physical properties. Alternatively, IDT homopolymers can be obtained via oxidative direct arylation polycondensation of IDT monomers (oxDAP), leading to IDT homopolymers with similar properties but at reduced yield. Detailed characterization by NMR, IR, UV-vis and PL spectroscopy, and thermal properties, is used to guide synthesis and to explain varying field-effect transistor hole mobilities in the range of $10^{-6}$- $10^{-3}$ cm$^2$/Vs.

INTRODUCTION

In recent years indacenodithiophene (IDT) based materials have been investigated extensively.[1–3] Both small molecules[4–7] and copolymers[1,8–11] have demonstrated high performance when used as active materials in organic photovoltaic (OPV)[12] devices and organic field-effect transistors (OFETs)[11]. The ladder-type five membered IDT moiety features desirable and easily tunable properties such as an extended π-conjugated system, an optimal balance of electron density and a quaternary bridging carbon atom that ensures planarity and carries different side groups for solubility and film morphology.[1,2] This coplanarity and the low energetic disorder of the electron rich fused ring aromatic structure favors π-electron delocalization and charge transport.[2,8,13–16] The attachment of side groups ensures high solubility for processing OPV[17,18] and OFET devices[5,8] from solution. Due to the major success of IDT-based materials, derivatization including backbone extension[19,20], different bridging atoms[21,22] and side chain variation[23,24] has been carried out.

Despite their success in terms of device performance, IDT-based materials generally suffer from tedious synthesis pathways. Several attempts to optimize the synthetic procedure aimed at minimizing and improving the reaction steps of monomer synthesis.[25–27] The methods of choice involve traditional transition-metal-catalyzed cross couplings such as Stille, Suzuki and Negishi reactions.[9,28,29] Despite being commonly established, each of these variants adhere to major disadvantages. For instance, Stille couplings produce toxic reagents and by-products and require additional purification steps. Negishi couplings used for IDT monomer synthesis rely on organometallic reagents of limited stability, and render the already lengthy reaction sequence towards IDT copolymers unnecessarily lengthy. C-H activation, also referred to as direct arylation (DA), is an atom-economically, highly attractive alternative.[30–32] The usage of simple C-H building blocks as monomers for direct arylation polycondensation (DAP) to furnish conjugated polymers brings about a reduced number of synthetic steps, streamlined protocols and opens up unprecedented possibilities in terms of molecular weight control.[31,33–37] While initially rather inefficient, recent protocols for high-yielding reactions that can be conducted at low catalyst loadings became available. To date, few IDT monomers have been used in DAP protocols.[8,25,35] However, IDT monomer synthesis still involves Stille or Negishi cross couplings. Furthermore, side reactions such as homocouplings are possible for IDT monomers used in DAP-based copolymerization.[8,25,35] Particularly for the case of IDT copolymers, the identification and characterization of generally ubiquitous homocouplings is very challenging due to limited information contained in broad NMR signals.[25]

Herein, we report on synthetic pathways to IDT-based polymers in which all backbone C-C coupling steps are achieved via DA. First, alternating copolymers **P(K-*alt*-T2)** comprising a symmetric, ketone-functionalized phenylene monomer (**K**) and 2,2'-bithiophene (**T2**) are synthesized via DAP in quantitative yield. Polymer analogous reductions to the polyalcohol (**PA**) and finally cyclization sequences are optimized and characterized in detail. At the same time, this method gives access to asymmetric substitution patterns at the bridging $sp^3$ carbon of the IDT unit, which may allow for tuning fundamental properties such as glass transition temperature ($T_g$), solubility as well as interchain interactions in pristine films and binary blends. The characteristics of these IDT polymers are finally compared to analogues made via oxidative direct C-H arylation polycondensation (oxDAP) of the IDT monomer.

RESULTS AND DISCUSSION

**Material synthesis and characterization.** The two synthetic routes towards IDT-based polymers are shown in Scheme 1. In the nomenclature of the polymers, A and B refer to polymer analogous cyclization and oxDAP, respectively.

**Scheme 1.** Synthetic routes towards **PIDT**. Route A: polymer analogous synthesis; Route B: synthesis of **IDT** monomer followed by polymerization via oxDAP.

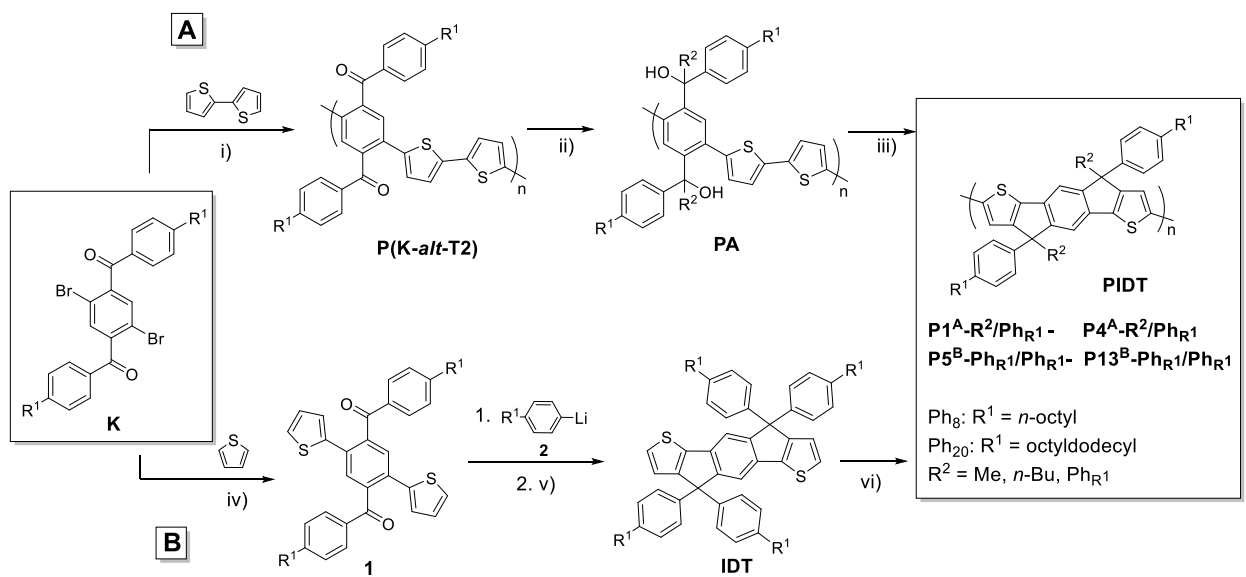

[a]Reaction conditions: (i) Pd$_2$dba$_3$ (1 mol%), P(o-anisyl)$_3$ (5 mol%), PivOH (1 eq), K$_2$CO$_3$ (3 eq), mesitylene (0.25 M), 100 °C, 72 h; (ii) R$^2$-Li (8 eq), toluene (4·10$^{-3}$ M), r.t., 3 h; (iii) BF$_3$·OEt$_2$ (18 eq), DCM (4·10$^{-3}$ M), r.t., 3 h; (iv) PCy$_3$ Pd G2 (5 mol%), PivOH (1 eq), K$_2$CO$_3$ (3 eq), THF (0.2 M), 100 °C; (v), Octane/AcOH (1:1, v/v, 0.02 M), H$_2$SO$_4$, 125 °C, 0.5 h, (vi) Pd(OAc)$_2$ (10 mol%), Cu(OAc)$_2$ (2.1 eq), K$_2$CO$_3$ (2.2 eq), DMAc (0.1 M), 110 °C, 72 h, air.

Copolymerization of **K** with 2,2'-bithiophene was first approached using the extraordinarily clean DAP conditions developed by Matsidik et al.[37] First polymerizations starting from n-octyl-substituted **K** led to polyketones with low solubility unsuitable for polymer analogous reactions. Instead, R$^1$ = 2-octyldodecyl was used, leading to larger molar mass and reaction yield. Due to the isostructural properties of **K** and 2,6-dibromonaphthalene diimide (NDIBr$_2$)[37], we envisioned transfer of these conditions for the synthesis of **P(K-*alt*-T2)** to occur without major optimization. However, to get **P(K-*alt*-T2)** in high yield, optimization of monomer concentration, temperature and catalyst/ligand loading was required, apparently due to a lower reactivity of **K** compared to NDIBr$_2$ (Table 1). Under optimized conditions, polyketones with molar masses $M_{n,SEC}$ up to 21 kg/mol were obtained quantitatively. Details of $^1$H and $^{13}$C NMR analyses are reported in the Supporting Information (Figure S1). The resulting polymers are highly soluble in common solvents as required for polymer analogous reactions. Polymers **P1$^A$-Me/Ph$_{20}$, P2$^A$-Bu/Ph$_{20}$, P3$^A$-**

**Ph$_{20}$/Ph$_{20}$** were finally synthesized as described by Scherf et al.[27] To obtain a clean reaction, the literature protocol had to be modified. Increased concentration and a reduced amount of the corresponding lithium organyls as well as reaction time afford **PIDT P1$^A$-Me/Ph$_{20}$, P2$^A$-Bu/Ph$_{20}$, P3$^A$-Ph$_{20}$/Ph$_{20}$** in 60-80% yield. The use of methyl lithium and n-butyl lithium led to an asymmetric substitution pattern in the corresponding homopolymers **P1$^A$-Me/Ph$_{20}$** and **P2$^A$-Bu/Ph$_{20}$**. Symmetric substitution could be achieved by transforming the synthesized 1-bromo-4-(2-octyldodecyl) benzene into the lithium analogue with n-butyl lithium. The progress of the post-polymerization sequence was monitored by IR and NMR spectroscopy. The reduction of the carbonyl group in the first reaction step can be well proved by disappearance of the stretching band at 1603 cm$^{-1}$ and 1665 cm$^{-1}$ (Figure 1). In case of the butyl derivative **P2$^A$-Bu/Ph$_{20}$**, the IR spectrum shows unreacted carbonyl groups (Figure 1, red box). A possible explanation is the higher reactivity of n-BuLi compared to e.g. MeLi, which opens up further reaction possibilities for the former such as reaction with THF or bithiophene end groups. Therefore, the temperature and equivalents of n-BuLi were varied. However, neither a decreased temperature nor the use of an excess of n-BuLi led to an improvement. All samples of the butyl derivative showed residual carbonyl bands in the IR spectrum and also a significant hypsochromic shift of the absorbance maximum in UV-vis spectra compared to **P1$^A$-Me/Ph$_{20}$** and **P3$^A$-Ph$_{20}$/Ph$_{20}$** (Figure S2).

**Table 1.** Summary of reaction conditions for the syntheses of **P(K-*alt*-T2)** via direct arylation polycondensation.[a]

| Entry | R$^1$ | Solv | Time (h)[b] | Conc (M) | T (°C) | Pd$_2$dba$_3$ (mol%) | P-(*o*-anisyl)$_3$ (mol%) | $M_n/M_w$ (kg/mol)[c] | Đ[c] | Yield (%)[d] |
|---|---|---|---|---|---|---|---|---|---|---|
| 1 | C8 | Tol | 72 | 0.5 | 100 | 1 | - | 6/11 | 1.8 | 33 |

| Entry | | | | | | | | | |
|---|---|---|---|---|---|---|---|---|---|
| 2 | C8 | Tol | 14 | 0.5 | 100 | 1 | - | 6/8 | 1.3 | 9 |
| 3 | C8 | Tol | 72 | 0.5 | 90 | 1 | - | 6/9 | 1.6 | 52 |
| 4 | C8 | Tol | 72 | 0.25 | 100 | 1 | - | 4/5 | 1.2 | 21 |
| 5 | C8 | Tol | 72 | 0.25 | 90 | 1 | - | 7/9 | 1.3 | 32 |
| 6 | C8 | Mes | 24 | 0.5 | 100 | 1 | - | 6/8 | 1.4 | 34 |
| 7 | C8 | Mes | 24 | 0.25 | 100 | 1 | - | 10/15 | 1.5 | 18 |
| 8[e] | C8 | Mes | 72 | 0.5 | 90 | 1 | - | - | - | - |
| 9 | C8 | Mes | 72 | 0.5 | 120 | 1 | - | 7/11 | 1.6 | 34 |
| 10 | C8 | Mes | 24 | 0.5 | 100 | 1 | 5 | 6/8 | 1.4 | 52 |
| 11 | C8 | Mes | 72 | 0.4 | 100 | 1 | 5 | 6/9 | 1.4 | 52 |
| 12 | C8 | Mes | 72 | 0.3 | 100 | 1 | 5 | 7/8 | 1.2 | 48 |
| 13 | C8 | Mes | 2 | 0.25 | 100 | 1 | 5 | 8/11 | 1.4 | 44 |
| 14 | C8 | Mes | 72 | 0.4 | 90 | 1 | 5 | 7/8 | 1.3 | 59 |
| 15 | C8 | Mes | 72 | 0.4 | 90 | 5 | 20 | 8/10 | 1.4 | 30 |
| 16 | C8C12 | Mes | 72 | 0.25 | 100 | 1 | 5 | 21/36 | 1.7 | 91 |
| 17 | C8C12 | Mes | 72 | 0.25 | 90 | 1 | 5 | 16/26 | 1.6 | 99 |
| 18 | C8C12 | Mes | 72 | 0.5 | 100 | 1 | 5 | 16/18 | 1.6 | 97 |
| 19 | C8C12 | Mes | 72 | 0.25 | 100 | 1 | 5 | 15/19 | 1.3 | 77 |
| 20[e,f] | C8C12 | Mes | 72 | 0.25 | 100 | 1 | 5 | - | - | - |
| 21[e] | C8C12 | Mes | 72 | 0.25 | 100 | 1 | - | - | - | - |

[a]1 eq PivOH and 3 eq K$_2$CO$_3$ were used in all entries. Tol and Mes are toluene and mesitylene, respectively. [b]Reduced reaction time in case of gelation. [c]From SEC in THF. [d]Isolated yield after Soxhlet extraction. [e]No chloroform fraction after Soxhlet extraction. [f]Slight excess (5 mol%) of ketone **K-Ph$_{20}$** was used.

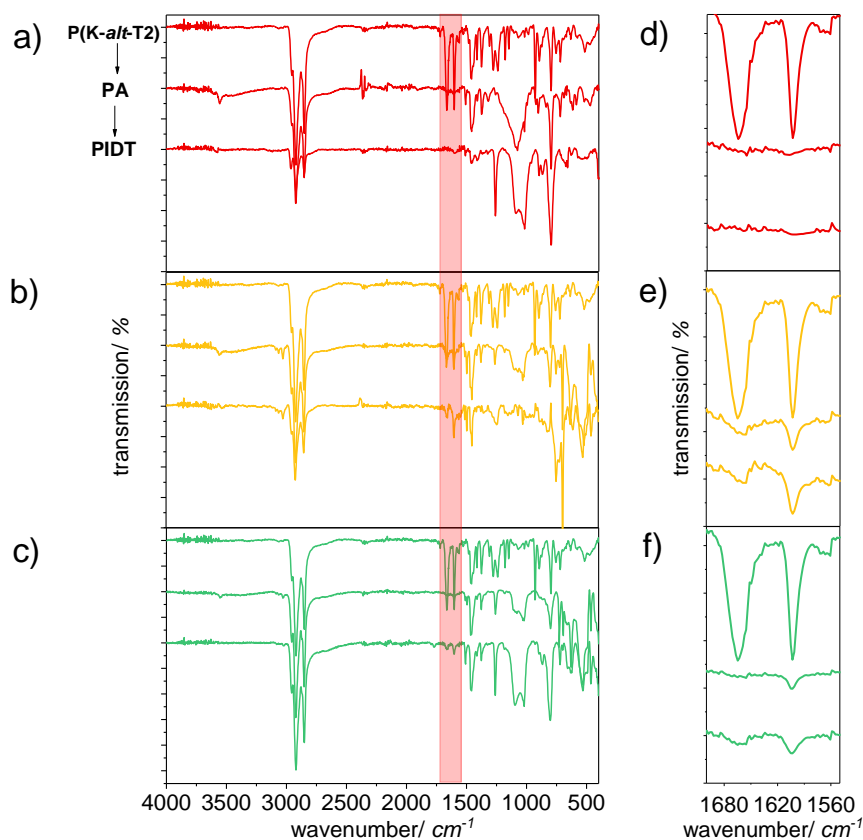

**Figure 1**. IR spectra of polymer-analogous conversion of **P(K-*alt*-T2)** to PIDTs **P1$^A$-Me/Ph$_{20}$** (a), **P2$^A$-Bu/Ph$_{20}$** (b), **P3$^A$-Ph$_{20}$/Ph$_{20}$** (c) including the spectra of the alcohol intermediates (**PA**). The region of carbonyl bands of the precursor **P(K-*alt*-T2)** is highlighted and enlarged in d) – f).

The SEC curves of **P1$^A$-Me/Ph$_{20}$** and **P2$^A$-Bu/Ph$_{20}$** displayed bimodal distributions, and a monomodal one for **P3$^A$-Ph$_{20}$/Ph$_{20}$** (Figure S3). Bimodal distributions are commonly associated with chain-chain coupling reactions. Interestingly, despite the bimodal distribution of **P1$^A$-Me/Ph$_{20}$**, clear indications for defects were not found by other characterization methods. One possibility is oxidative chain-chain coupling of two thiophene chain ends, which may be addressed by endcapping the polyketone with bromobenzene as further described below.

Detailed NMR signal assignments of IDT copolymers are rare in the literature due to the low content of information contained in broad signals of $^1$H NMR spectra. To assist in signal assignment and eventually reveal defect structures, NMR spectroscopy was aided by model compounds **8** and **9** (Supporting Information, Figure S4 and S5). Cyclization can be well proven by $^{13}$C NMR spectra, where the signal of the keto group at $\delta$ = 197 ppm vanished and the quaternary carbon shows up in the 52 – 64 ppm region depending on $R^2$ (Figure S5). The different substitution pattern of the quaternary ring carbon ($C_8$) can be traced by its $^{13}$C chemical shift (52.7 ppm for Me/alkylphenyl (**8**) vs. 57.1 ppm for *n*-Bu/alkylphenyl (**9**) vs. 62.7 ppm for 4-alkylphenyl/4-alkylphenyl (**IDT-Ph$_{20}$**)). Very similar chemical shifts were observed for $C_8$ in **P1$^A$-Me/Ph$_{20}$** – **P3$^A$-Ph$_{20}$/Ph$_{20}$** proving successful cyclization (Figure 2). For **P3$^A$-Ph$_{20}$/Ph$_{20}$** the appearance of a major signal at 63.1 ppm ($R^2$ = 4-(2-octyldodecyl)phenyl) and a minor signal at 57.4 ppm ($R^2$ = *n*-butyl) points to a side reaction (Figure 2c). Because the added 4-alkylphenyl lithium was obtained from the bromide and *n*-butyl lithium without further purification, unreacted reactant led to a mixture of 4-octyldodecylphenyl and *n*-butyl substitution in the final cyclized polymer. For **P1$^A$-Me/Ph$_{20}$** and **P2$^A$-Bu/Ph$_{20}$**, four different substituents at both $C_8$ result in two diastereomers for the repeat unit as proved by signal splitting for model compounds **8** and **9**. Since both $R^2$ can be on the same (*R,R/S,S*-racemate) or different sides (*R,S-meso-* isomer) of the IDT π-plane, they lead to regioirregular backbone structures.

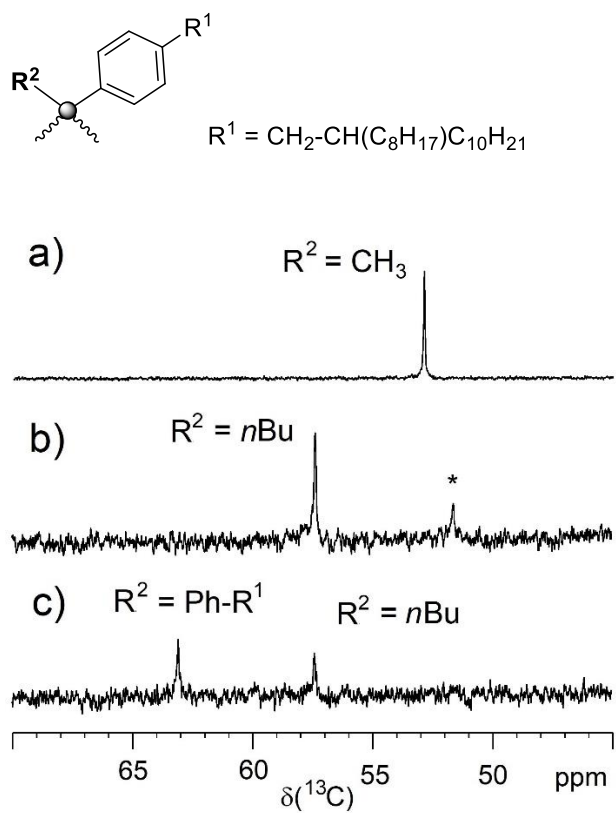

**Figure 2.** Region of the quaternary carbon signal of the IDT unit of **PIDT** polymers **P1$^A$-Me/Ph$_{20}$** (a), **P2$^A$-Bu/Ph$_{20}$** (b) and **P3$^A$-Ph$_{20}$/Ph$_{20}$** (c) in CDCl$_3$. * marks an impurity.

Whereas the signals of the aliphatic moieties can be well observed both in the $^1$H and $^{13}$C NMR spectra of the polymers, only **P1$^A$-Me/Ph$_{20}$** gives well resolved spectra for the IDT moiety (Figure S6a and S7a). The $^1$H and $^{13}$C NMR spectrum fully confirm the desired **PIDT** structure and do not point to structural defects. However, the origin of signal broadening observed for **P2$^A$-Bu/Ph$_{20}$** and **P3$^A$-Ph$_{20}$/Ph$_{20}$** with bulkier R$^2$ residue is unclear. Possibly, the *n*-butyl and 4-alkylphenyl groups decrease segmental backbone motion whereas side-chain motion is less influenced. Because the spin-spin relaxation times (T$_2$) are sensitive to changes in molecular motions they also influence linewidth ($\Delta\nu_{1/2} \sim 1/T_2$). Thus, a decreasing T$_2$ value for the backbone nuclei could cause the line broadening. Apart from the broad signals, the chemical shifts are in accordance with those of **P1$^A$-Me/Ph$_{20}$**.

**Optical and thermal properties.** The formation of the planar rigid **PIDT** structure can be followed by UV-vis spectroscopy (Figure 3a). The change in conformation and also of the electronic properties of the polymer backbone leads to major shifts of the absorption maximum. Compared to the polyketones, the polyalcohols (**PA1$^A$-Me/Ph$_{20}$, PA2$^A$-Bu/Ph$_{20}$, PA3$^A$-Ph$_{20}$/Ph$_{20}$**) exhibit blue-shifted absorption bands by ~77 nm, which is caused by torsion and eventually by a diminished push-pull character of the backbone. In turn, cyclization increases conjugation and thus leads to a bathochromic effect. A closer look into the optical properties of the synthesized homopolymers exposes significant differences. First discrepancies can already be observed in the optical spectra of the polyalcohols. The absorption bands of **PA2$^A$-Bu/Ph$_{20}$** and **PA3$^A$-Ph$_{20}$/Ph$_{20}$** are broadened compared to **PA1$^A$-Me/Ph$_{20}$**. This behavior is also reflected in the cyclized polymers. **P1$^A$-Me/Ph$_{20}$** shows a structured absorption band with maximum at 515 nm and a well resolved vibronic side band at 544 nm (Figure 3). **P3$^A$-Ph$_{20}$/Ph$_{20}$** displays a similar shape and absorption maximum with a slightly less distinct shoulder. The resolution of this vibronic structure - caused by the rigid structure of the IDT backbone - can also be seen in the UV-vis spectra of the model compounds **8**, **9** and **IDT-Ph$_{20}$**. In all cases the absorption maximum is around 355 nm with a sharp and well resolved side band at around 374 nm (Figure S8). In comparison, the band of **P2$^A$-Bu/Ph$_{20}$** shows no sign of a side band. But it is obviously broadened with a less steep absorption edge and a significantly blue-shifted maximum. The emission spectra of the model compounds **8**, **9** and **IDT-Ph$_{20}$** show a blue-shift of the maximum with increasing size of the second side chain $R^2$ (Figure S8). This effect cannot be observed in the emission spectra of the corresponding homopolymers **P1$^A$-Me/Ph$_{20}$, P3$^A$-Ph$_{20}$/Ph$_{20}$** and **P2$^A$-Bu/Ph$_{20}$** (Figure 3b). In general, a band with an emission maximum around 565 nm and a shoulder at around 610 nm is visible. The shape of **P2$^A$-Bu/Ph$_{20}$** is also broadened and exhibits a bigger Stokes-shift of $\Delta \lambda_{Stokes}$ = 87 nm than **P1$^A$-Me/Ph$_{20}$** and **P3$^A$-Ph$_{20}$/Ph$_{20}$** ($\Delta \lambda_{Stokes} \approx 55$ nm).

The anomaly of **P2$^A$-Bu/Ph$_{20}$** may be caused by incomplete conversion of the carbonyl groups. A possible side reaction may be lithiation of –T2 end groups of **P(K-*alt*-T2)** by BuLi followed by an interchain reaction with a carbonyl group. In an attempt to prove the appearance of such defect structures, a model reaction was carried out (Scheme S1). The model reaction demonstrates that lithiated T2 is able to attack the carbonyl group of compound **K** leading to model compound **10** (Scheme S1). The structure could be verified by NMR spectroscopy (Figure S9). Such defects would lead to branching. In an additional control experiment, the polymer precursor **P(K-*alt*-T2)** was terminated with bromobenzene to eliminate –T2 end groups. The NMR analysis of **P(K-*alt*-T2)$_{term}$** confirms the successful termination (Figure S10). During the following polymer analogous cyclization leading to **P2$^A_{term}$-Bu/Ph$_{20}$**, possible chain-chain coupling should be suppressed. However, characterization by NMR spectroscopy did not reveal significant differences compared to **P2$^A$-Bu/Ph$_{20}$** without phenyl termination (Figures S11 and S12). UV-vis and IR spectra display the same discrepancy as observed for **P2$^A$-Bu/Ph$_{20}$** (Figure S13a and Figure S14). Only in the molecular weight distribution of **P2$^A_{term}$-Bu/Ph$_{20}$**, a high molecular weight shoulder was no longer seen (Figure S13b). From these results we conclude that cyclization by using BuLi is disadvantageous as side reactions, likely branching, are involved. Such side reactions would explain the hypsochromic shift in the UV-vis spectrum and also the lower solubility of **P2$^A$-Bu/Ph$_{20}$**. But also the very electron rich T2 end groups may undergo chain chain coupling as an additional side reaction, which can simply detected by SEC but is hard to prove directly by NMR spectroscopy (and not at all by IR and UV-vis spectroscopy). However, such chain chain coupling will cause quaterthiophene main chain defects, which pose an electronic alteration to the backbone structure. This aspect will be taken up again when discussing field-effect transistor characterization.

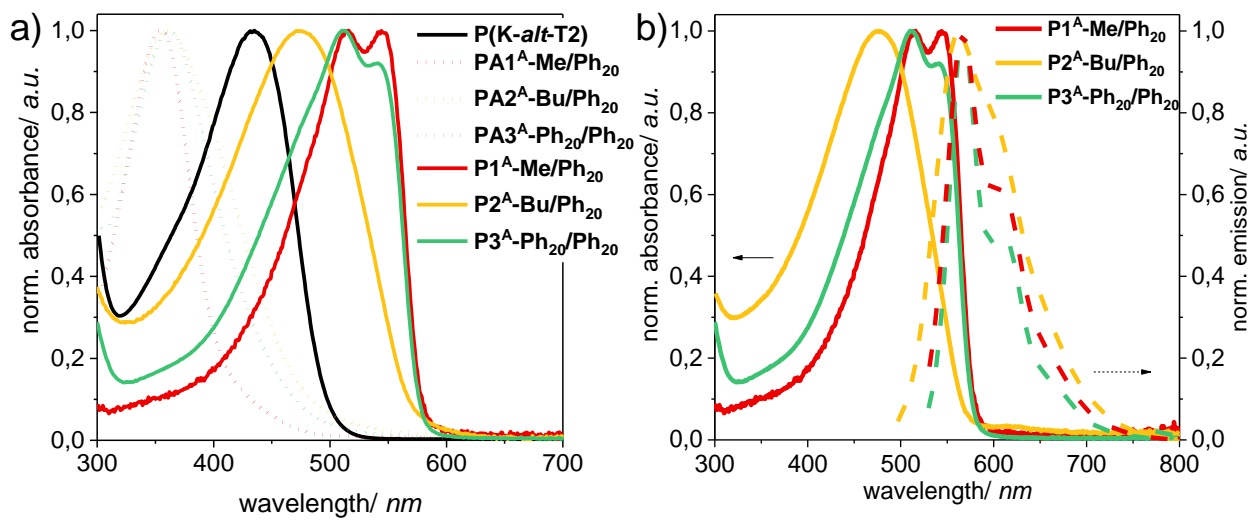

**Figure 3.** a) Reaction control by UV-vis spectroscopy of polymer analogous conversion to **PIDTs** in CHCl$_3$ solution at r.t (**PA**: crude product after conversion with lithium compound). b) Absorption and emission spectra of homopolymers in CHCl$_3$ solution at r.t.

Polymers **P1$^A$-Me/Ph$_{20}$** and **P2$^A$-Bu/Ph$_{20}$** show good thermal stability with degradation temperatures higher than 350 °C and 250 °C, respectively (Figure S15). In contrast, polymer **P3$^A$-Ph$_{20}$/Ph$_{20}$** shows two onset temperatures. The first appears at about 100 °C with a weight loss less than 5 % and the second is higher than 350 °C with a weight loss of about 15 % (Figure S15). The thermal properties were also examined by differential scanning calorimetry (DSC). The **PIDTs** show glass transitions in between 40 – 70 °C (Figure S16). Because of the quaternary carbons with asymmetric, bulky substituents ordered packing is more difficult leading to amorphous nature.[2,19,28,29,38,39] All characteristic data are summarized in Table 2.

**Table 2.** Molecular weights, optical and thermal properties of synthesized polymers.

| Entry | Monomer | R$^2$ | $M_n/M_w$ (kg/mol)$^a$ | Đ$^a$ | $\lambda_{max,abs}$ (nm)$^b$ | $\lambda_{max,em}$ (nm)$^b$ | $T_g$ (°C) | Yield (%)$^c$ |
|---|---|---|---|---|---|---|---|---|

| Polymer | Monomer | R | $M_n/M_w$ [kg/mol][a] | Đ[a] | Abs λ_max [nm][b] | Em λ_max [nm][b] | Φ [%] | Yield [%][c] |
|---|---|---|---|---|---|---|---|---|
| P(K-*alt*-T2), entry 19 | K-Ph$_{20}$ | - | 15/19 | 1.3 | 435 | | | 99 |
| P1$^A$-Me/Ph$_{20}$ | K-Ph$_{20}$ | Me | 30/80 | 2.7 | *516*/544 | *568*/614 | 59 | 58 |
| P2$^A$-Bu/Ph$_{20}$ | K-Ph$_{20}$ | *n*-Bu | 20/44 | 2.2 | 477 | 564 | 42 | 72 |
| P3$^A$-Ph$_{20}$/Ph$_{20}$ | K-Ph$_{20}$ | Ph$_{20}$ | 18/22 | 2.0 | *510*/541 | *567*/612 | 65 | 64 |
| P4$^A$-Ph$_8$/Ph$_8$[d] | K-Ph$_8$ | Ph$_8$ | 5/11 | 2.3 | 496 | 566 | - | 50 |
| P11$^B$-Ph$_{20}$/Ph$_{20}$ | IDT-Ph$_{20}$ | Ph$_{20}$ | 16/18 | 1.1 | 510 | *568*/603 | 59 | 4 |
| P13$^B$-Ph$_8$/Ph$_8$ | IDT-Ph$_8$ | Ph$_8$ | 17/22 | 1.3 | *514*/538 | *569*/611 | 64 | 6 |

[a]From SEC in THF, [b]In CHCl$_3$ solution at r.t., [c]Overall yield starting from monomer **K**. Maxima with the highest intensity in italic type, [d]Due to low molecular weight and small amount of sample material no further characterization was performed.

**IDT homopolymers via oxidative direct arylation polycondensation and comparison with polymer analogous pathway.** In the second approach to IDT-based polymers, the fused monomer **IDT** was synthesized first followed by homopolymerization via oxidative direct arylation (oxDAP).[39–41] As starting material for **IDT** monomer compound **K** was chosen instead of the usually used ester derivative.[25] Compound **1** was made via DA with thiophene (Scheme 1b) in 48% yield after the optimization of catalyst/ligand, solvent and concentration. In comparison to route A, the limiting aspect of route B is not solubility but synthesis and purification of monomer **IDT**. Especially purification of **IDT-Ph$_{20}$** by column chromatography was challenging and accompanied by low yields. Due to the long and branched side chains, the reactivity of intermediates was lower and the final product was obtained as oil. Usage of *n*-octyl resulted in **IDT-Ph$_8$** being a solid, which facilitated synthesis and purification. The additional possibility to recrystallize the product after

column chromatography lead to higher purity as needed for polymerization and yields up to 53 %. Due to only few literature protocols on oxDAP[40,42,43], optimization of catalyst loading, atmosphere and oxidant was required (Table 3).

The best results could be achieved with 20 mol% of Pd catalyst and copper acetate as oxidant (Table 3, entry **P8$^B$-Ph$_{20}$/Ph$_{20}$**). Remarkably, no inert gas atmosphere is necessary. Hence, it can be assumed that oxygen from air may act as co-oxidant. Nevertheless, the molecular weights achieved were low in all cases. Values comparable to route A could only be realized after repeated polymerization of an oligomer fraction under the same conditions (Table 3, entries **P10$^B$-Ph$_{20}$/Ph$_{20}$/P11$^B$-Ph$_{20}$/Ph$_{20}$/P13$^B$-Ph$_8$/Ph$_8$**, Figure S17).

**Table 3.** Summary of reaction conditions and molecular weights of **PIDT** via oxDAP.$^a$

| Entry | Pd(OAc)$_2$ (mol%) | Oxidant[a] | Base[a] | Atmosphere | $M_n/M_w$ (kg/mol)[b] | $Đ$[b] | Yield (%)[c] |
|---|---|---|---|---|---|---|---|
| **P5$^B$-Ph$_{20}$/Ph$_{20}$** | 10 | Cu(OAc)$_2$ | K$_2$CO$_3$ | argon | 6/7 | 1.2 | 23[e] |
| **P6$^B$-Ph$_{20}$/Ph$_{20}$** | 10 | Cu(OAc)$_2$ | K$_2$CO$_3$ | air | 8/11 | 1.3 | 31 |
| **P7$^B$-Ph$_{20}$/Ph$_{20}$** | 10 | Ag$_2$CO$_3$ | K$_2$CO$_3$ | argon | 7/8 | 1.2 | 12[e] |
| **P8$^B$-Ph$_{20}$/Ph$_{20}$** | 20 | Cu(OAc)$_2$ | K$_2$CO$_3$ | air | 12/15 | 1.2 | 39 |
| **P9$^B$-Ph$_{20}$/Ph$_{20}$**[d] | 2 | Ag$_2$CO$_3$ | K$_2$CO$_3$/ AcOH | argon | - |  | - |
| **P10$^B$-Ph$_{20}$/Ph$_{20}$**[e] | 10 | Cu(OAc)$_2$ | K$_2$CO$_3$ | air | 24/35 | 1.5 | 22 |
| **P11$^B$-Ph$_{20}$/Ph$_{20}$**[e] | 20 | Cu(OAc)$_2$ | K$_2$CO$_3$ | air | 16/18 | 1.1 | 40 |
| **P12$^B$-Ph$_8$/Ph$_8$** | 10 | Cu(OAc)$_2$ | K$_2$CO$_3$ | air | 16/29 | 1.8 | 38 |
| **P13$^B$-Ph$_8$/Ph$_8$**[e] | 10 | Cu(OAc)$_2$ | K$_2$CO$_3$ | air | 17/22 | 1.3 | 10 |

[a]2.1 eq oxidant, 2.2 eq additives and DMAc as solvent (0.1 M) were used in all entries, [b]From SEC in THF, [c]Isolated yield after Soxhlet extraction with acetone, ethyl acetate (EA) and chloroform, [d]Only acetone soluble material of low molecular weight (Mn ~ 2 kg/mol), [e]Polymerization of EA soluble fraction of **P6$^B$-Ph$_{20}$/Ph$_{20}$/P8$^B$-Ph$_{20}$/Ph$_{20}$/P12$^B$-Ph$_8$/Ph$_8$** under same conditions.

The resulting polymers were further characterized by NMR, IR, UV-vis spectroscopy, photoluminescence, TGA as well as DSC analysis and compared to their polymer analogues from route A. NMR spectra of **P11$^A$-Ph$_{20}$/Ph$_{20}$** are provided in the Supporting Information (Figure S6d and Figure S18a). As expected the $^{13}$C-NMR spectrum shows only one signal of the quaternary carbon at 63.0 ppm. In general, the chemical shifts are similar to the polymer analogous synthesized material **P3$^A$-Ph$_{20}$/Ph$_{20}$**. Due to the signal broadening no information about structural defects or end groups could be extracted.

The IR spectra display similar characteristics in all cases (Figure 4). The weak band between 1600 cm$^{-1}$ and 1700 cm$^{-1}$ is usually an indication of the appearance of carbonyl groups. In fact, in case of **P3$^A$-Ph$_{20}$/Ph$_{20}$** a C=O stretching band would be a possible explanation. Due to the polymer analogous reaction pathway (route A) the first assumption would be an incomplete ring closure with residual carbonyl groups of the precursor polymer **P(K-*alt*-T2)** being still present. However, the comparison of the IR spectra with those of the monomer **IDT-Ph$_{20}$** and the polymers **P11$^B$-Ph$_{20}$/Ph$_{20}$** and **P13$^B$-Ph$_8$/Ph$_8$** made via route B, reveals the same weak band at 1600 cm$^{-1}$. As carbonyl groups are not involved here, the weak band between 1600 cm$^{-1}$ and 1700 cm$^{-1}$ may alternatively caused by the two bulky aromatic side chains. Aromatic C=C stretching bands show weak signals in this region as well.[44] Therefore, we assume that in case of **P3$^A$-Ph$_{20}$/Ph$_{20}$** this band can also be attributed to an aromatic C=C stretching and not to C=O stretching by incomplete ring closure.

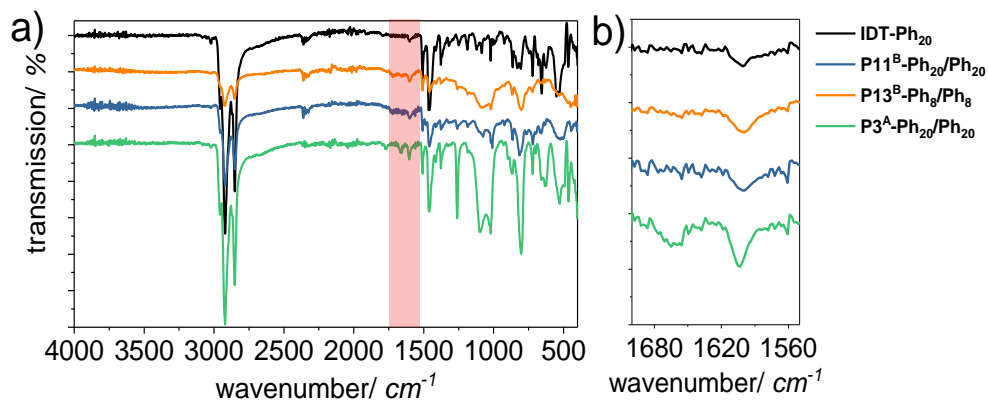

**Figure 4.** a) Comparison of IR spectra of PIDTs via route A (**P3$^A$-Ph$_{20}$/Ph$_{20}$**) and route B (**P11$^B$-Ph$_{20}$/Ph$_{20}$**, **P13$^B$-Ph$_8$/Ph$_8$**). b) shows the enlarged wavenumber range of the red box.

In contrast, the characterization by optical methods shows significant differences (Figure 5). In case of **P4$^A$-Ph$_8$/Ph$_8$** ($M_n$ = 5 kg/mol) and **P13$^B$-Ph$_8$/Ph$_8$** ($M_n$ = 17 kg/mol) the varied maxima and shapes of the spectra mainly result from different conjugation lengths with the low molecular weight of **P4$^A$-Ph$_8$/Ph$_8$** leading to blue shifted absorption maximum. In contrast, **P3$^A$-Ph$_{20}$/Ph$_{20}$** ($M_n$ = 18 kg/mol) and **P11$^B$-Ph$_{20}$/Ph$_{20}$** ($M_n$ = 16 kg/mol) display similar optical properties. Nevertheless, **P3$^A$-Ph$_{20}$/Ph$_{20}$** shows a sharper and better resolved absorption as well as emission than **P11$^B$-Ph$_{20}$/Ph$_{20}$**. The broadened and undefined bands of **P11$^B$-Ph$_{20}$/Ph$_{20}$** may be attributed to impurities caused by tedious monomer purification or structural defects.

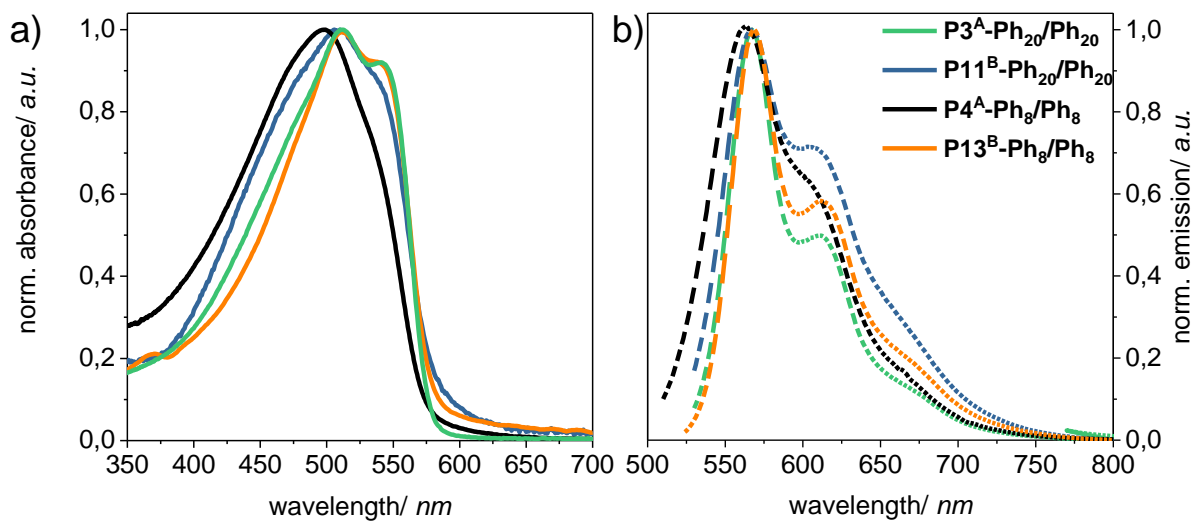

**Figure 5.** Comparison of UV-vis (a) and emission spectra (b) of **PIDT**s via route A (**P3$^A$-Ph$_{20}$/Ph$_{20}$, P4$^A$-Ph$_8$/Ph$_8$**) and route B (**P11$^B$-Ph$_{20}$/Ph$_{20}$, P13$^B$-Ph$_8$/Ph$_8$**) in CHCl$_3$ solution at r.t.

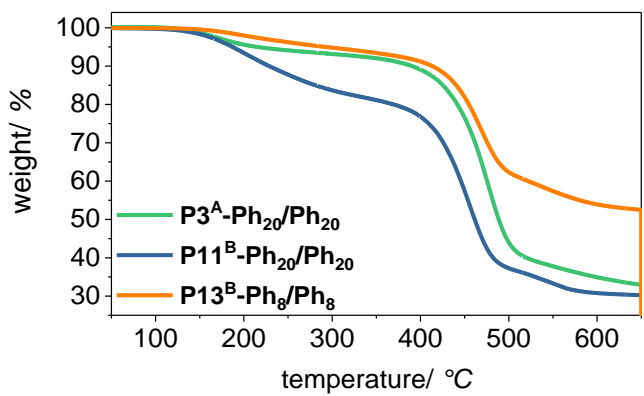

**Figure 6.** Comparison of TGA thermograms of PIDTs via route A (**P3$^A$-Ph$_{20}$/Ph$_{20}$**) and route B (**P11$^B$-Ph$_{20}$/Ph$_{20}$, P13$^B$-Ph$_8$/Ph$_8$**) in N$_2$.

This also explains the 100 °C lower degradation temperature of **P11$^B$-Ph$_{20}$/Ph$_{20}$** (Figure 6). The best thermal stability shows **P13$^B$-Ph$_8$/Ph$_8$** with its 4-octylphenyl substitution pattern. The homopolymers were further investigated by DSC measurements. Glass transition temperatures of **P11$^B$-Ph$_{20}$/Ph$_{20}$** and **P13$^B$-Ph$_8$/Ph$_8$** can be seen at 59 °C and 64 °C, respectively (Table 2). As expected for the amorphous morphologies of IDT-based polymers, crystallization and melting temperatures were absent (Figure S16). Interestingly, the side group mixture of 4-octyldodecylphenyl and *n*-butyl substitution in **P3$^A$-Ph$_{20}$/Ph$_{20}$** has no significant influence on the optical and thermal properties.

To investigate the electronic properties of the IDT copolymers made by different routes, organic field effect transistors (OFETs) were fabricated. All devices show p-type behaviour. **P1$^A$-Me/Ph$_{20}$** and **P3$^A$-Ph$_{20}$/Ph$_{20}$** display ideal transfer characteristics with similar source-to-drain currents and field-effect mobilities in the order of $10^{-4}$-$10^{-3}$ cm$^2$/Vs. These two samples exhibit the least structural defects according to IR spectroscopy and show the most structured UV-vis absorption bands. The highest mobility of **P3$^A$-Ph$_{20}$/Ph$_{20}$** may further be rationalized by its monomodal SEC curve indicating the absence of chain chain couplings. Assuming quaterthiophene main chain defects as a result of T2 chain end coupling as the reason for the bimodal SEC curves of **P1$^A$-Me/Ph$_{20}$** and **P2$^A$-Bu/Ph$_{20}$**, this electronic main chain defect would explanation their lower electronic performance and at the same time the superior mobility of **P3$^A$-Ph$_{20}$/Ph$_{20}$.** Clearly, **P2$^A$-Bu/Ph$_{20}$** with clear evidence of chain chain coupling, incomplete cyclization or branching shows only moderate current modulation and poor field-effect mobility. These spectroscopic properties of **P13$^B$-Ph$_8$/Ph$_8$** are comparable to the better defined samples from route A, but hole mobilities are in the $10^{-5}$ cm$^2$/Vs range only. The origin of this discrepancy is unclear, especially considering the similar molecular weights of **P3$^A$-Ph$_{20}$/Ph$_{20}$** and **P13$^B$-Ph$_8$/Ph$_8$**. However, most of the difference comes from a much larger threshold voltage, indicating either poorer injection at the dielectric-semiconductor interface. A summary of hole mobilities extracted from the saturation regime ($V_d$ = -60 V) is reported in Figure 7.

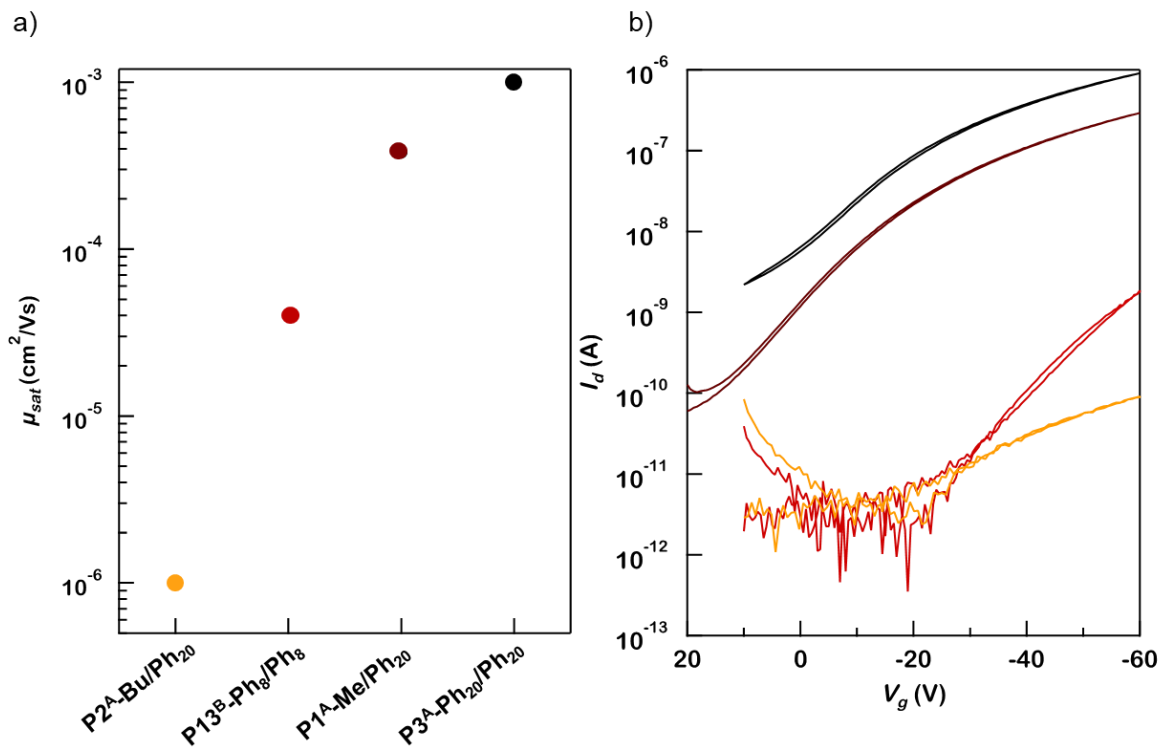

**Figure 7.** a) Field-effect hole mobilities extracted from the saturation regime. b) Representative transfer characteristic curves measured in saturation regime, $V_d$ = -60 V.

CONCLUSIONS

We introduced and optimized two synthetic ways to IDT-based homopolymers using exclusively C-H activation for C-C bond formation. Pathway A includes direct arylation polycondensation (DAP) for polyketone synthesis followed by reduction to the polyalcohols and cyclization. In pathway B, oxidative direct arylation polycondensation (oxDAP) is used. Using pathway A, polyketones with satisfying molecular weight up to $M_{n,SEC}$= 21 kg/mol were obtained in quantitative yield. Only methyl lithium and lithiated phenyl derivatives as reagents for cyclization led to complete conversion and well-defined polymers with asymmetric substitution pattern, while butyl lithium caused side reactions such as chain chain coupling and/ or branching. The $^1$H and $^{13}$C

NMR spectra of IDT homopolymers with methyl groups exhibited significantly sharper signals, making analysis and conclusions more straightforward, despite a concomitant lower solubility. Potentially, pathway A can be transferred to thiophene-acceptor-thiophene comonomers to yield IDT copolymers that are known for their excellent performance in solar cells and field effect transistor devices. Pathway B involving oxDAP led to IDT homopolymers of similar spectroscopic characteristics but at significantly lower yield. The best charge carrier mobilities up to $10^{-3}$ cm$^2$/Vs were obtained from IDT homopolymers made by pathway A and showing monomodal SEC curves. Generally, the determination of structural defects with high precision is most challenging for this class of conjugated polymers, but the herein presented characterization may serve as a starting point for both simplified synthesis protocols as well as yet more detailed and advanced characterization.

EXPERIMENTAL SECTION

**Materials.** All starting materials were purchased from commercial sources and used without further purification unless otherwise specified. 2, 2'-Bithiophene (T2) was obtained from Alfa Aesar (98%) and further purified by eluting through a silica plug with petroleum ether (PE). All reactions were carried out in flame dried glassware and under dry inert gas atmosphere. The detailed syntheses of compound **K** and model compounds **8** and **9** are shown in Scheme S2 and Scheme S3 in the Supporting Information (SI). They were prepared as reported.[45–47]

*Monomer syntheses.* Compounds **1** and **2** were synthesized as detailed in the SI. Monomer **IDT** was synthesized according to literature procedures.[25,41]

*General route to **P(K-alt-T2)***. Compound **K** (133.7 μmol, 1 eq), 2,2'-bithiophene (22.2 mg, 133.7 μmol, 1 eq), pivalic acid (13.7 mg, 133.7 μmol, 1 eq), potassium carbonate (55.4 mg, 400.9 μmol, 3 eq) were placed in a vial and dissolved in 0.5 mL degassed mesitylene. Then Pd$_2$dba$_3$ (1.2 mg, 1 mol%) and P(*o*-anisyl)$_3$ (2.4 mg, 5 mol%) were added under argon and stirred for 72 h at 100 °C. After cooling to room temperature, the mixture was diluted with chloroform, precipitated into methanol and purified by Soxhlet extraction with acetone, ethyl acetate and chloroform. The chloroform fraction was filtered through a silica gel plug to afford an orange solid.

*General route to IDT homopolymers via route A (**P1$^A$-Me/Ph$_{20}$, P2$^A$-Bu/Ph$_{20}$, P3$^A$-Ph$_{20}$/Ph$_{20}$**).* To a solution of **P(K-*alt*-T2)** (entry 19) (50 mg, 49.4 μmol, 1 eq) in 10 mL toluene at room temperature the corresponding lithium compound (395.4 μmol, 8 eq) was added and after 30 minutes 3 mL THF was added. After stirring for 3 h at room temperature, the reaction mixture was quenched with ethanol and water, extracted with chloroform and dried over magnesium sulfate. The solvent was removed under vacuum and the crude product was immediately dissolved in dry dichloromethane. After the addition of boron trifluoride diethyl etherate (0.1 mL, 130.3 mg, 938.9 μmol, 19 eq) the mixture was stirred for 3 h at room temperature and then quenched with ethanol and water, extracted with chloroform and dried over magnesium sulfate. The crude product was precipitated into methanol and purified by Soxhlet extraction with acetone, ethyl acetate and chloroform. The chloroform fraction was filtered through a silica gel plug to afford **P1$^A$-Me/Ph$_{20}$, P2$^A$-Bu/Ph$_{20}$, P3$^A$-Ph$_{20}$/Ph$_{20}$** as red solid.

*General route to IDT homopolymers via route B (**P5$^B$-Ph$_{20}$/Ph$_{20}$-P13$^B$-Ph$_8$/Ph$_8$**).* Compound **IDT** (21.3 μmol, 1 eq), potassium carbonate (6.5 mg, 46.8 μmol, 2.2 eq) and copper(II) acetate (8.1 mg, 44.7 μmol, 2.1 eq) were placed into a vial and dissolved in dry *N*,*N*-dimethylacetamide. Then Pd(OAc)$_2$ (0.5 mg, 10 mol%) was added and the mixture was stirred for 72 h at 110 °C under air. After cooling to room temperature, the mixture was diluted with chloroform, precipitated into methanol and purified by Soxhlet extraction with acetone, ethyl acetate and chloroform. The chloroform fraction was filtered through a silica gel plug to afford **P5$^B$-Ph$_{20}$/Ph$_{20}$-P13$^B$-Ph$_8$/Ph$_8$** as red solids.

**Supporting Information**. Additional data including details of synthesis, measurements as well as characterization of model compounds; additional NMR, SEC, IR, UV-vis and DSC data are provided.

**Author information**

**Corresponding Author**

* <michael.sommer@chemie.tu-chemnitz.de>

**Author Contributions**

The manuscript was written through contributions of all authors. All authors have given approval to the final version of the manuscript.

**Acknowledgment**

The authors acknowledge M. Raisch and S. Schmidt for DSC measurements, J. Streif and D. Stegerer for TGA measurements, and V. Pacheco for initial NMR measurements. Funding from DFG (SO 1213/8-1) is greatly acknowledged. A. S. and M. C. acknowledge the financial support of the European Research Council, European Union's Horizon 2020 Research and Innovation Programme HEROIC under Grant 638059.

**Conflict of interest** The authors declare that they have no conflict of interest.